\documentclass[a4paper,fleqn,usenatbib]{mnras}

\usepackage[utf8]{inputenc}
\usepackage{newtxtext,newtxmath}

\usepackage[T1]{fontenc}
\usepackage{ae,aecompl}

\usepackage{graphicx}	
\usepackage{amsmath}	

\newcommand{\msun}{\mbox{$M_{\odot}$}}
\newcommand{\Msun}{\mbox{$M_{\odot}$}}
\newcommand{\lsun}{\mbox{$L_{\odot}$}}

\newcommand{\zsun}{\mbox{$Z_{\odot}$}}
\newcommand{\teff}{\mbox{$T_{\rm eff}$}}
\newcommand{\Teff}{\mbox{$T_{\rm eff}$}}

\newcommand{\vinf}{\mbox{$\varv_{\infty}$}}
\newcommand{\vesc}{\mbox{$\varv_{\rm esc}$}}

\newcommand{\mdot}{\mbox{$\dot{M}$}}

\newcommand{\msunyr}{\mbox{$M_{\odot} {\rm yr}^{-1}$}}

\def\aap{A\&A}                
\def\ao{Applied Optics}       
\def\apj{ApJ}                 
\def\apjl{ApJ}                
\def\apjs{ApJS}               
\def\araa{ARA\&A}             
\def\mnras{MNRAS}             
\title[Wind parameters at low metallicity]{Metallicity-dependent wind parameter predictions for OB stars}

\author[J.S. Vink \& A.A.C. Sander]{Jorick S. Vink,$^1$\thanks{E-mail: jorick.vink@armagh.ac.uk}
and 
Andreas A. C. Sander$^1$
\\
$^1$ Armagh Observatory and Planetarium, College Hill, Armagh, BT61 9DG, Northern Ireland, UK\\
}

\date{Accepted 2021 March 20. Received 2021 March 20; in original form 2020 September 02}

\pubyear{2020}

\begin{document}
\label{firstpage}
\pagerange{\pageref{firstpage}--\pageref{lastpage}}
\maketitle

\begin{abstract}
Mass-loss rates and terminal wind velocities are key parameters that determine the kinetic wind energy and momenta of massive
  stars. Furthermore, accurate mass-loss rates
  determine the mass and rotational velocity evolution of mass stars, and their fates as neutron stars and black holes in function of metallicity ($Z$).
  Here we update our Monte Carlo mass-loss Recipe with new dynamically-consistent
  computations of the terminal wind velocity -- as a function of $Z$. These predictions are particularly timely as the
  HST ULLYSES project will observe ultraviolet spectra with blue-shifted P Cygni lines of hundreds of massive stars in the low-$Z$ Large and Small Magellanic Clouds, as well as sub-SMC metallicity hosts. Around 35\,000 K, we uncover a weak-wind "dip" and we present diagnostics
  to investigate its physics with ULLYSES and X-Shooter data. We discuss how the dip may provide important information on wind-driving physics, and how this is of key relevance towards finding a new gold-standard for OB star mass-loss rates.
  For B supergiants below the Fe {\sc iv} to {\sc iii} bi-stability jump, the terminal
  velocity is found to be independent of $Z$ and $M$, while the mass-loss rate still varies as $\dot{M} \propto Z^{0.85}$. For O-type stars above the bi-stability
  jump we find a terminal-velocity dependence of $\vinf \propto Z^{0.19}$ and the $Z$-dependence of the mass-loss rate is
  found to be as shallow as $\dot{M} \propto Z^{0.42}$, implying that to reproduce the `heavy' black holes from LIGO/Virgo, the `low $Z$' requirement becomes
  even more stringent than was previously anticipated. 
\end{abstract} 



\section{Introduction}
\label{s_intro}

Massive OB-type stars are dominant forces of Nature: they shine at high luminosities, sculpt the interstellar medium (ISM) 
through strong line-driven winds, ending their lives with powerful supernovae or silent collapses into black holes (e.g. Langer 2012). 
The outcome of their evolution is largely a function of their stellar wind mass-loss rate, which is a function of host galaxy 
metal content, $Z$. The puzzle of the fate of massive stars as a function of $Z$ has become more pertinent now that gravitational
waves (GWs) of merging black holes (BHs) have indicated the existence of "heavy" BHs, up to 30-40\msun (Abbott et al. 2016), and since recently even up to 85\msun (Abbott et al. 2020; Vink et al. 2020).
The most logical reason for the very heavy Nature of these BHs would be the occurrence of GW events in low-$Z$ host galaxies, as a natural consequence of reduced Fe-dependent wind mass loss 
(Vink \& de Koter 2005; Eldridge \& Vink 2006; Belczynski et al. 2010). 

The radiation-driven wind theory (RDWT) was first developed by Lucy \& Solomon (1970) and Castor et al. (1975; CAK), before it underwent
significant updates of the line opacities in the 1980s (Abbott 1982; Friend \& Abbott 1986; Pauldrach et al. 1986), resulting in a modified CAK theory that still relied on a parameterised line acceleration as a function of location in the wind. 
According to Puls et al. (2008) and Martins \& Palacios (2013) most current day stellar evolution and population synthesis models employ a newer version of the RDWT where the line acceleration is computed at each position in the wind, in the form of the Vink et al. (2000) recipe
which is based on the Monte Carlo radiative transfer approach from Abbott \& Lucy (1985). The original Monte Carlo models were semi-empirical in that they assumed 
terminal wind velocities (\vinf), which are relatively straightforwardly available from observed P\,Cygni line profiles in the ultraviolet (UV) part
of the electro-magnetic spectrum. Thanks to the unprecedented Hubble Space Telescope (HST) Ultraviolet Legacy Library of Young Stars as Essential Standards
(ULLYSES) there will soon be a flurry of activity in determining new terminal wind velocities and mass-loss rates of hundreds of
massive OB stars in the low-$Z$ environments of the Small \& Large Magellanic Clouds, and even at sub-SMC metallicities. For this reason, an improved set of model predictions for not
only the wind mass-loss rate, but also the terminal wind velocity is needed, and 
we will employ the locally consistent approach of M\"uller \& Vink (2008) to predict 
velocity structures and mass-loss rates for a range of OB supergiants, as well as their metallicity ($Z$) dependence, which we compare to our earlier Vink et al. (2001) mass-loss slope.

In Sect.~\ref{s_model}, we briefly describe the Monte Carlo modelling and our   
physical assumptions. In Sect.~\ref{s_res} mass-loss rates and wind terminal
velocities are presented for 20, 30, and 60 \msun\ stars across the temperature regime of the 
bi-stability jump. We present UV wind diagnostics of stars around the 35\,000 K weak wind `dip' in Sect.\,4, before we discuss \& conclude.


\section{Physical assumptions and Monte Carlo modelling}
\label{s_model}

In this paper we simultaneously predict mass-loss rates (\mdot) and wind velocity 
Parameters (\vinf, $\beta$) based on the M\"uller \& Vink (2008) approach (see also Muijres et al. 2012).
The underlying model atmosphere is the Improved Sobolev Approximation code {\sc isa-wind} by 
de Koter et al. (1993) where the basic effects of the diffuse radiation field are included in 
line resonance zones. In the models of this paper we compute H, He, C, N, O, S, Si, explicitly in non-LTE, as we only found minor 
differences when treating Fe in the modified nebular approximation (Schmutz 1991).
 
One of the strengths of the code as well as alternative co-moving frame (CMF) codes such as {\sc cmfgen} (Hillier \& Miller 1998), {\sc PoWR} (Sander et al. 2020) and {\sc fastwind} (Puls et al. 2005),
is that it treats the star (`core') and the wind (`halo') in a unified manner, i.e. 
without a core-halo configuration.
In the Monte Carlo radiative transfer, lines are described using the Sobolev approximation, which 
is an excellent approximation in the outer portions of stellar winds, where velocity gradients are substantial.
This may provide confidence in our aim of predicting the outer wind dynamics and terminal wind velocity 
accurately.  However, if subtle non-Sobolev effects in the inner parts of the wind are relevant, this
could have relevant implications for the predicted mass-loss values, as CMF calculations 
seem to provide lower mass-loss rates by a factor of a few 
(Petrov et al. 2016; Krticka \& Kubat 2017; Sander et al. 2017; Sundqvist et al. 2019). However, it remains
  yet unknown if these sophisticated CMF computations also provide reliable absolute mass-loss rates.
	Empirical determinations of $\dot{M}$ (e.g.\ Bouret et al. 2013; Ramachandran et al. 2018; Bestenlehner et al. 2020) so 
	far yield a complex picture with some results being in agreement with Monte Carlo predictions, while others yield lower
	mass-loss rates.

  At least for the most massive transitional Of/WNh stars, we know that the Sobolev-based Monte Carlo models
  provide model independent and accurate results, independent of clumping and porosity uncertainties (Vink \& Gr\"afener 2012) that undermine the accuracy of empirical mass-loss rates at lower mass.
  Future empirical tests will need to provide answers to the issue of the absolute values of
  mass-loss rates in the canonical O-star regime studied here.

Observational and theoretical line transitions have been adopted from Kurucz as previously
(Kurucz \& Bell 1995). The abundances are from Anders \& Grevesse (1989).
While the effects of a smaller solar metallicity (Asplund et al. 2009)
could lead to lower mass-loss predictions, given the dominance of Fe in fixing the mass-loss rate, 
the differences may turn out to be relatively minor. The main motivation for keeping the 
abundances the same as in previous 
(Vink et al. 1999; 2000; 2001)
works is that it enables direct comparisons.

The Monte Carlo models are both homogeneous on small scales and spherically symmetric on larger scales. Whilst 
wind clumping results in downward readjustments 
of empirical mass-loss rates by factors 
of about 3 (Hillier 1991; Moffat \& Robert 1994; Davies et al. 2007; Puls et al. 2008; 
Hamann et al. 2008; Sundqvist et al. 2014; Ram\'irez-Agudelo et al. 2017), it should be realised that it 
may affect the theoretical line driving in the opposite direction, unless the medium becomes extremely porous 
(Muijres et al. 2011; Jiang et al. 2015; Shaviv 2000; Owocki 2015), with associated implications for spectral features 
(Oskinova et al. 2007; Surlan et al. 2013; Sundqvist et al. 2014; Petrov et al. 2014).

\section{Dynamically consistent Results}
\label{s_res}

We compute mass-loss rates as a function of effective temperature and $Z$ for a series of 
stellar masses. In the Vink et al. (2000, 2001) studies, we computed a large grid of 
12 model series covering a range of masses and luminosities. It was found that the prime factor setting the mass-loss rates for different masses was 
related to the luminosity-to-mass {\it ratio}, or alternatively to the Eddington parameter, $\Gamma_{\rm e}$. 
For models in close proximity to the Eddington limit we refer the reader to Vink et al. (2011) and Vink (2018a;b). Here we focus on the more canonical
stellar masses relating to Eddington modest factors in the range 0.1 - 0.4. 
We pick three characteristic masses of 20, 30, and 60 solar masses, 
corresponding to Eddington parameters of 0.13, 0.27, and 0.43. 
Luminosities were those as in model series \#3, \#6, and \#10 from Vink et al. (2000) for direct comparison. 

For the dynamically consistent modelling we employ a convergence criterium where the terminal wind velocities derived from Eqs.\,(14) and\,(15) from Muijres et al. (2012) are within 10\% from one another (see also M\"uller \& Vink 2008).

\subsection{Wind parameters for all 3 masses at solar metallicity}

\begin{figure}
\includegraphics[width=\columnwidth]{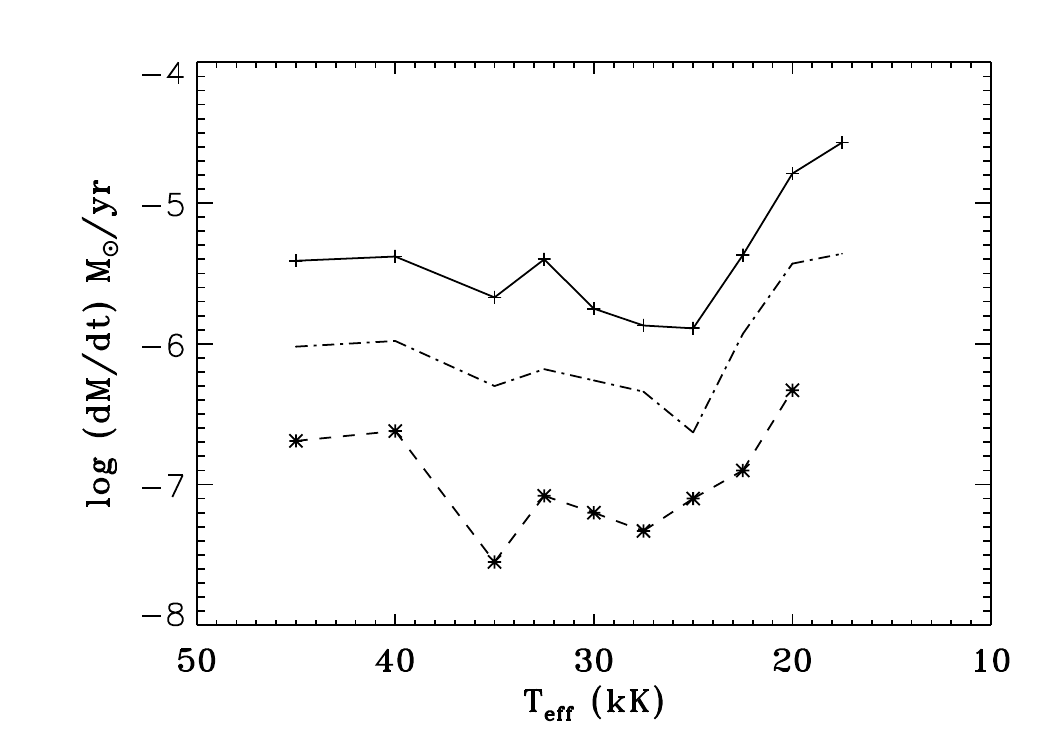}
\caption{The predicted mass-loss rates ($\mdot$) versus \Teff\ for all 3 model series at \zsun. The solid line represents the highest $M = 60\msun$ model, the lowest mass $M = 20\msun$ gives the lowest 
Mass-loss rates (dashed line) whilst the dashed-dotted intermediate values are for the $M = 30\msun$ model.}
\label{f_mdot}
\end{figure}

\begin{figure}
\includegraphics[width=\columnwidth]{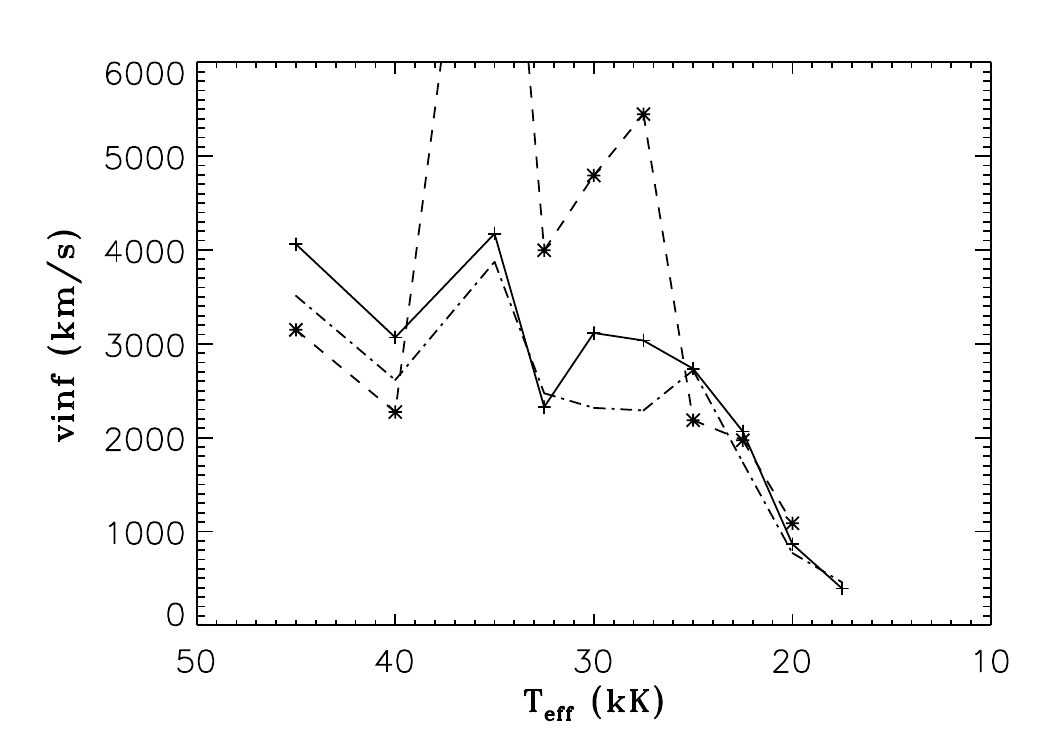}
\caption{The predicted wind velocity ($\vinf$) versus \teff\ for all 3 model series at \zsun. The solid line represents the highest $M = 60\msun$ model, the lowest mass $M = 20\msun$ gives the lowest 
Mass-loss rates (dashed line) whilst the dashed-dotted intermediate values are for the $M = 30\msun$ model.}
\label{f_vinf}
\end{figure}

Tables\,\ref{tab:results1} to \ref{tab:results3} list the locally-consistent Monte Carlo predictions for
all 3 model series. The stellar parameters are identical to model series \#6, \#10 and \#3 of Vink et al. (2000, 2001). 
They are listed in columns (1) - (4), whilst 
predicted wind terminal velocities, new mass-loss rates, and 
wind acceleration parameter $\beta$ from $\varv(r) = \vinf (1 - r/R)^\beta$ are 
listed in columns (5) - (7). 
The mass-loss predictions for solar metallicity are shown in Fig.\,\ref{f_mdot}, and 
the predicted terminal wind velocities are displayed in Fig.\,\ref{f_vinf}.

In agreement with initial results of Vink (2018b), Figure 1 shows that the mass-loss rates 
increase drastically -- by an order of magnitude -- around 21\,000 K. This is the so-called bi-stability jump. 
The terminal wind velocity is found to drop significantly over the same 
temperature range (Fig.\,2). 
The newly computed mass-loss rates are in excellent
agreement with the Vink et al. (2000, 2001) rates, to within the 0.08 dex mass loss uncertainty of the Vink 00/01 recipe,
{\it when accounting for the newly predicted terminal wind velocities}. This should be no surprise as, although the hydrodynamical treatment is different, the
  radiative transfer method is similar (see also Muijres et al. 2012).

Figure 2 shows relatively low terminal wind velocities on the cool side of the bi-stability jump 
(down to 300 km/s) and values in the range 2000-4000 km/s for hotter objects. 
Whilst some of the drop in wind velocity is due to the fact that the stellar 
escape velocity also drops at lower \teff\ (cf.\,Fig.4), due to the larger radii, the {\it abrupt} drop at the jump temperature 
is due to the change in the ionisation from Fe {\sc iv} to Fe {\sc iii} (Vink et al. 1999). 
Interestingly, on the cool side of the bi-stability jump, the terminal wind velocities are found to be independent of 
stellar mass (and as we will see later, also independent of $Z$). On the hot side of the jump, there is capricious behaviour for predicted terminal wind velocities.
This is mainly due to the fact that in a relative sense CNO lines are more important than Fe at these higher \teff\ values (Vink et al. 1999),
and as there are far fewer lines of CNO than Fe, the effects of individual CNO lines become noticeable for the \vinf\ determination (see also Krticka \& Kubat 2017; Sander et al. 2018; Bj{\"o}rklund et al. 2020). 
The 20\msun\ model exhibits particularly dramatic behaviour around 35\,000 K which we attribute to the same issues as those that 
Muijres et al. (2012) experienced for lower mass/luminosity models. This was attributed
to the onset of the `weak-wind problem' (Martins et al. 2005; Puls et al. 2008; Marcolino et al. 2009; de Almeida et al. 2019), that might be related to the `inverse' bi-stability effect
due to a lack of Fe {\sc iv} driving around spectral type O6.5 -- corresponding to $\log(L/\lsun) = 5.2$ (Muijres et al. 2012).

\subsection{Wind parameters as a function of $Z$ for the canonical 30\msun\ model}

Table\,\ref{tab:results1} lists the locally-consistent Monte Carlo predictions for our canonical
30\,$\msun$ star for 5 metallicities.
The mass-loss predictions for all $Z$ are shown in Fig.\,3, and 
the predicted terminal wind velocities are displayed in Fig.\,4.
Figure 3 shows that the mass-loss bi-stability jump shifts to lower \teff\ values for lower $Z$ in qualitative and quantitative agreement 
with Vink et al. (2001). The reason is that for lower wind densities at lower $Z$ the Fe {\sc iv} to Fe {\sc iii} recombination occurs at lower \teff.

Figure 4 shows the same capricious wind velocity behaviour for low $Z$ models as was present at the 20\msun\ model at solar $Z$. 
Interestingly, Fig.\,4 demonstrates
that the wind terminal velocity is $Z$-dependent for models on the hot side of the bi-stability jump, but it is $Z$ {\it independent} for models {\it below} the jump.
This implies that the way the wind parameters of mass-loss rate and wind velocity, will undergo some changes with respect to the simpler
wind momentum scaling of $\mdot \times \vinf\ \propto Z^{0.85}$ that was found universally in the global models of Vink et al. (2001). 

\begin{figure}
\includegraphics[width=\columnwidth]{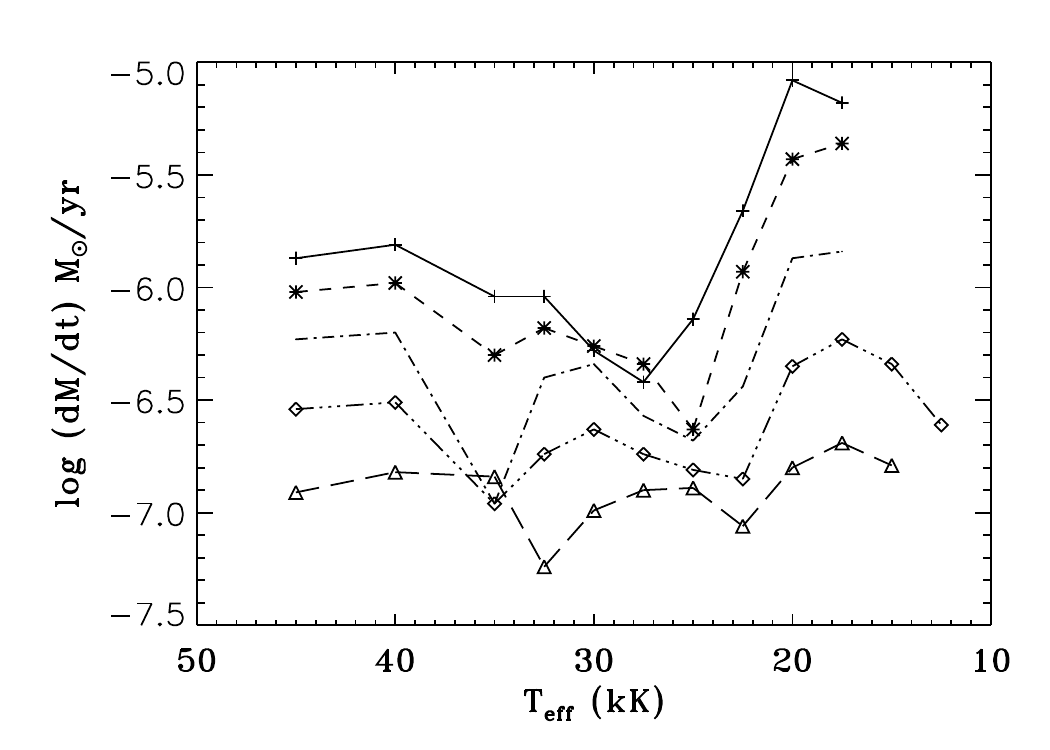}
\caption{The predicted mass-loss rates ($\mdot$) versus \Teff\ for the canonical 30\msun\ model. The solid line is for 3\zsun, whilst the dashed, dashed-dotted, dashed-dot-dot-dotted, and long-dashed lines are for \zsun, a third \zsun, a tenth \zsun, and 1 over 33 \zsun, respectively.}
\label{f_mdotM30}
\end{figure}

\begin{figure}
\includegraphics[width=\columnwidth]{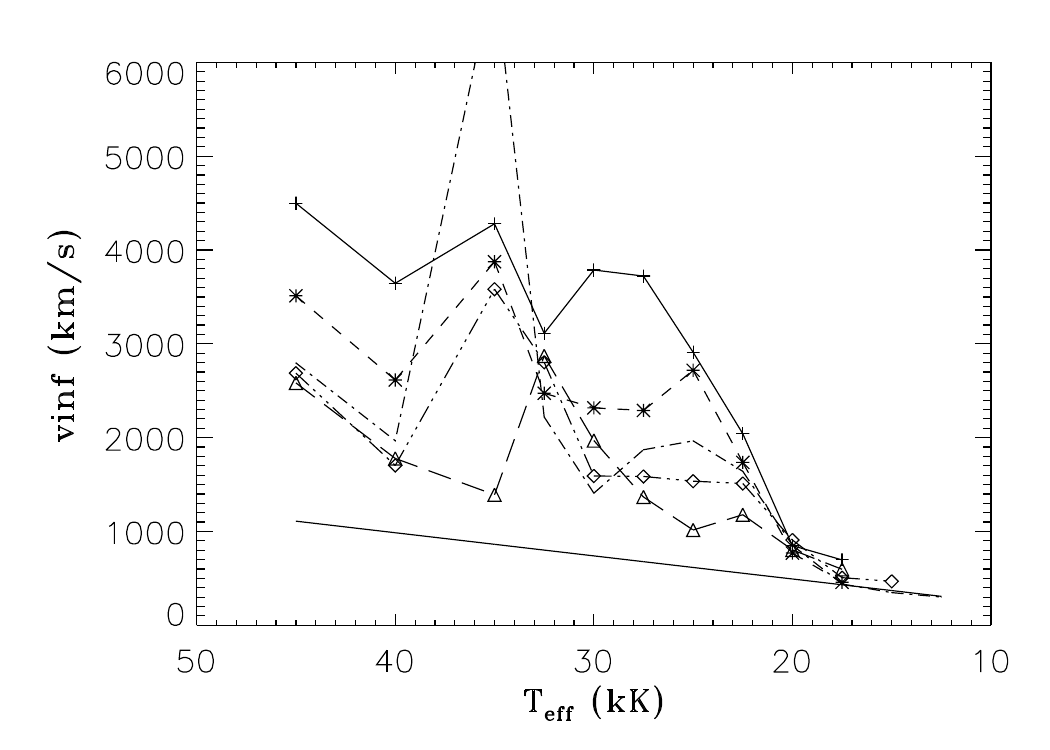}
\caption{The predicted wind velocity ($\vinf$) versus \teff\ for the canonical 30\msun\ model. The solid line is for 3\zsun, whilst the dashed, dashed-dotted, dashed-dot-dot-dotted, and long-dashed lines are for \zsun, a third \zsun, a tenth \zsun, and 1 over 33 \zsun, respectively. The line at the bottom denotes the stellar escape velocity as a function of \teff.}
\label{f_vinfM30}
\end{figure}

\begin{table}
\centering
\begin{tabular}{llllccc}
\hline
\hline
$M$      & $\log L$ & \teff  & $Z/\zsun$ & \vinf  & $\log \mdot$ & \(\beta\)\\
(\Msun)  &  (\lsun)  &  (kK) &           & (km/s) &   (\msunyr) & \\  
\hline
30      & 5.5     & 45     & 3          & 4499          & $-$5.87     & 1.04\\
        &         &        & 1          & 3512          & $-$6.02     & 0.96\\
        &         &        & 1/3        & 2796          & $-$6.23     & 0.85\\
        &         &        & 1/10       & 2687          & $-$6.54     & 0.77\\
        &         &        & 1/33       & 2583          & $-$6.91     & 0.85\\[0.5em]

        &         & 40     & 3          & 3645          & $-$5.81     & 0.98\\  
        &         &        & 1          & 2614          & $-$5.98     & 0.81\\
        &         &        & 1/3        & 1967          & $-$6.20     & 0.78\\
        &         &        & 1/10       & 1703          & $-$6.51     & 0.69\\
        &         &        & 1/33       & 1778          & $-$6.82     & 0.69\\[0.5em] 

        &         & 35     & 3          & 4279          & $-$6.04     & 1.22\\
        &         &        & 1          & 3875          & $-$6.30     & 1.10\\
        &         &        & 1/3        & 6970          & $-$6.96     & 1.74\\
        &         &        & 1/10       & 3582          & $-$6.96     & 1.18\\
        &         &        & 1/33       & 1391          & $-$6.84     & 0.79\\[0.5em]

        &         & 32.5  & 3           & 3112          & $-$6.04     & 1.07\\
        &         &       & 1           & 2473          & $-$6.18     & 1.03\\
        &         &       & 1/3         & 2220          & $-$6.40     & 0.84\\
        &         &       & 1/10        & 2803          & $-$6.74     & 0.97\\
        &         &       & 1/33        & 2870          & $-$7.24     & 0.99\\[0.5em]

        &         & 30    & 3           & 3787          & $-$6.28    & 1.13\\
        &         &       & 1           & 2317          & $-$6.26    & 1.01\\
        &         &       & 1/3         & 1408          & $-$6.34    & 0.74\\
        &         &       & 1/10        & 1591          & $-$6.63    & 0.74\\
        &         &       & 1/33        & 1969          & $-$6.99   & 0.78\\[0.5em]

        &         & 27.5  & 3           & 3722          & $-$6.42  & 1.36\\
        &         &       & 1           & 2290          & $-$6.34  & 1.22\\
        &         &       & 1/3         & 1871          & $-$6.57  & 0.84\\
        &         &       & 1/10        & 1586          & $-$6.74  & 0.73\\
        &         &       & 1/33        & 1367          & $-$6.90  & 0.68\\[0.5em]

        &         & 25    & 3           & 2910          & $-$6.14  & 1.37\\
        &         &       & 1           & 2718          & $-$6.63  & 1.14\\
        &         &       & 1/3         & 1966          & $-$6.68  & 0.92\\
        &         &       & 1/10        & 1536          & $-$6.81  & 0.77\\
        &         &       & 1/33        & 1015          & $-$6.89  & 0.65\\[0.5em]

        &         & 22.5  & 3           & 2041          & $-$5.66  & 1.09\\
        &         &       & 1           & 1736          & $-$5.93  & 0.97\\
        &         &       & 1/3         & 1641          & $-$6.44   & 1.06\\
        &         &       & 1/10        & 1511          & $-$6.85  & 0.97\\
        &         &       & 1/33        & 1178          & $-$7.06  & 0.77\\[0.5em]

        &         & 20    & 3           &  849          & $-$5.08  & 0.84\\
        &         &       & 1           &  770          & $-$5.43   & 0.78\\
        &         &       & 1/3         &  853          & $-$5.87    & 0.75\\
        &         &       & 1/10        &  909          & $-$6.35    & 0.82\\
        &         &       & 1/33        &  804          & $-$6.80    & 0.78\\[0.5em]

        &         & 17.5  & 3           & 697           & $-$5.18   & 0.89\\
        &         &       & 1           & 458           & $-$5.36    & 0.67\\
        &         &       & 1/3         & 440           & $-$5.84   & 0.60\\
        &         &       & 1/10        & 507           & $-$6.23   & 0.65\\
        &         &       & 1/33        & 597          & $-$6.69   & 0.71\\[0.5em]

        &         & 15    & 1/10        & 348            & $-$6.34  & 0.63\\
        &         &       & 1/33        & 468            & $-$6.79  & 0.71\\[0.5em]

        &         & 12.5  & 1/10        & 301            & $-$6.61 & 0.98\\
\hline
\hline
\end{tabular}
\caption{Wind predictions for a 30 \msun\ model with stellar parameters identical
to model series \#6 from Vink et al. (2000,2001). Note that the terminal wind velocities only have a 10\% precision, so the large number of digits should not be interpreted as a precision indicator.}
\label{tab:results1}
\end{table}

\subsection{Wind parameters for the higher (60\,\msun) and lower (20\,\msun) mass models}

Table\,\ref{tab:results2} and\,\ref{tab:results3} list the locally-consistent Monte Carlo predictions for both
the higher mass (60\,$\msun$) and lower mass (20\,\msun) models -- for all 5 metallicities.
The mass-loss rate and wind velocity predictions for all $Z$ are shown in Figs.\,5-8.
Again, the mass-loss bi-stability jump shifts to lower \teff\ values for lower $Z$, and the wind terminal velocity is 
$Z$-dependent for models on the hot side of the bi-stability jump, but independent from $Z$ for models below the jump.

Combining the results for all masses, we find that for $T_{\rm eff} \ge 25\,000$K (hotter than the BS-jump):

\begin{equation}
\log \mdot \propto  1.48(\pm 0.04) \log L + 1.28(\pm 0.22) \log \teff + 0.42(\pm 0.03) \log Z        
\label{eq:fitMdHot}
\end{equation}

Note that the reason to not provide an explicit new mass-loss prescription but only proportionalities is that in terms of the $(M,L)$ grid, Vink et al. (2000) is the 
more extensive set of computations in terms of parameter space. In that work, $M$ and $L$ were varied independently from each other, and independent $M$ and an $L$ mass-loss 
dependencies were derived.

For $T_{\rm eff} \le 20\,000$K (cooler than the BS-jump), we find:

\begin{equation}
\log \mdot \propto  1.60(\pm 0.05) \log L + 0.6(\pm 0.38) \log \teff + 0.85(\pm 0.03) \log Z        
\label{eq:fitMdCool}
\end{equation}
Interestingly the \mdot\ vs. $Z$ dependence for the cool models is exactly the same as that for the wind momentum of
all global models computed in Vink et al. (2001), whilst the \mdot\ vs. $Z$ dependence for the hot models is much shallower than for the cool models.
The error bars on the \mdot\ vs. $Z$ dependencies is 0.03 for both temperature regimes.

For the wind terminal wind velocity (in km/s), we find (for 87 data-points, and $\sigma = 0.06$)
for $T_{\rm eff} \ge 25\,000$K (i.e. hotter than the BS-jump):

\begin{align}
\label{eq:fitVinfHot}
\log \vinf =  0.39(\pm 0.76) &- 0.04(\pm 0.03) \log(L/\lsun)\\
  \nonumber                  &+ 0.74(\pm 0.16) \log \teff + 0.19(\pm 0.02) \log(Z/\zsun)        
\end{align}
Whilst for for $T_{\rm eff} \le 20\,000$K (cooler than the BS-jump), we find (on the basis of 37 data-points, and $\sigma = 0.14$):

\begin{align}
\label{eq:fitVinfCool}
\log \vinf = -7.79(\pm 1.17) &- 0.07(\pm 0.04) \log(L/\lsun)\\
  \nonumber                  &+ 2.57(\pm 0.27) \log \teff -0.003(\pm 0.023) \log(Z/\zsun)       
\end{align}
In other words for cool models the wind terminal velocity is basically {\it in}dependent on $Z$, whilst 
the dependence of the wind velocity with $Z$ as $\vinf \propto Z^{0.19}$ for the hot models is slightly steeper than the $\vinf \propto Z^{0.13}$ relation computed by 
Leitherer et al. (1992) as was considered in Vink et al. (2001).
We do not provide wind terminal velocities in terms of stellar escape velocities, as the CAK scaling where $\vinf/\vesc \propto \sqrt{\alpha/(1-\alpha)}$ has limited value as the CAK parameter $\alpha$ is not constant with location through the wind (Vink 2000; Kudritzki 2002, Muijres et al. 2012). However, if readers are interested in comparing observed terminal velocities in terms of the escape velocity for comparison with both (modified) CAK theory (e.g. Howarth \& Prinja 1989; Lamers et al. 1995) as well as our newer RDWT predictions, the escape velocities can be obtained from the data in the Tables.

\begin{table}
\centering
\begin{tabular}{llllccc}
\hline
\hline
$M$      & $\log L$ & \teff  & $Z/\zsun$ & \vinf  & $\log \mdot$ & \(\beta\)\\
(\Msun)  &  (\lsun)  &  (kK) &           & (km/s) &   (\msunyr) & \\
\hline
60     & 6.0      & 45     & 3          & 5155          & $-$5.29     & 1.10\\
       &          &        & 1          & 4062          & $-$5.41     & 1.03\\
       &          &        & 1/3        & 3127          & $-$5.59     & 0.90\\
       &          &        & 1/10       & 2782          & $-$5.87     & 0.81\\
       &          &        & 1/33       & 2492          & $-$6.11     & 0.80\\[0.5em]

        &         & 40     & 3          & 4281          & $-$5.20     & 1.25\\
        &         &        & 1          & 3068          & $-$5.38     & 0.88\\
        &         &        & 1/3        & 2236          & $-$5.54     & 0.83\\
        &         &        & 1/10       & 1970          & $-$5.85     & 0.79\\
        &         &        & 1/33       & 1921          & $-$6.15     & 0.72\\[0.5em] 

        &         & 35     & 3          & 5120          & $-$5.51     & 1.32\\
        &         &        & 1          & 4174          & $-$5.67     & 1.25\\
        &         &        & 1/3        & 2715          & $-$5.79     & 0.99\\
        &         &        & 1/10       & 1483          & $-$5.84     & 0.73\\
        &         &        & 1/33       & 1051          & $-$6.12     & 0.64\\[0.5em]

        &         & 32.5  & 3           & 3429          & $-$5.37     & 1.17\\
        &         &       & 1           & 2327          & $-$5.40     & 0.98\\
        &         &       & 1/3         & 2522          & $-$5.75     & 0.98\\
        &         &       & 1/10        & 2496          & $-$6.02     & 0.96\\
        &         &       & 1/33        & 3178          & $-$6.51     & 1.05\\[0.5em]

        &         & 30    & 3           & 3214          & $-$5.45    & 1.29\\
        &         &       & 1           & 3115          & $-$5.75    & 1.14\\
        &         &       & 1/3         & 1680          & $-$5.69    & 0.86\\
        &         &       & 1/10        & 1280          & $-$5.86    & 0.74\\
        &         &       & 1/33        & 1421          & $-$6.16    & 0.74\\[0.5em]

        &         & 27.5  & 3           & 3831          & $-$5.71  & 1.38\\
        &         &       & 1           & 3035          & $-$5.87  & 1.24\\
        &         &       & 1/3         & 2174          & $-$5.98  & 0.96\\
        &         &       & 1/10        & 1471          & $-$6.07  & 0.77\\
        &         &       & 1/33        & 1437          & $-$6.28  & 0.74\\[0.5em]

        &         & 25    & 3           & 2968          & $-$5.52  & 1.49\\
        &         &       & 1           & 2735          & $-$5.89  & 1.24\\
        &         &       & 1/3         & 2185          & $-$6.09  & 1.05\\
        &         &       & 1/10        & 1690          & $-$6.23  & 0.83\\
        &         &       & 1/33        & 1337          & $-$6.35  & 0.72\\[0.5em]

        &         & 22.5  & 3           & 2285          & $-$5.14  & 1.10\\
        &         &       & 1           & 2066          & $-$5.37  & 1.09\\
        &         &       & 1/3         & 1884          & $-$5.83  & 1.13\\
        &         &       & 1/10        & 1651          & $-$6.25  & 1.01\\
        &         &       & 1/33        & 1272          & $-$6.42  & 0.84\\[0.5em]

        &         & 20    & 3           & 1132          & $-$4.60  & 0.85\\
        &         &       & 1           &  865          & $-$4.79   & 0.83\\
        &         &       & 1/3         &  883          & $-$5.20    & 0.77\\
        &         &       & 1/10        &  912          & $-$5.63    & 0.78\\
        &         &       & 1/33        &  960          & $-$6.14    & 0.85\\[0.5em]

        &         & 17.5  & 3           & 478           & $-$4.26   & 0.72\\
        &         &       & 1           & 392           & $-$4.57    & 0.65\\
        &         &       & 1/3         & 315           & $-$4.95   & 0.74\\
        &         &       & 1/10        & 459           & $-$5.50   & 0.65\\
        &         &       & 1/33        & 564          & $-$5.95   & 0.71\\[0.5em]

        &         & 15    & 1/10        & 366            & $-$5.64  & 0.70\\
        &         &       & 1/33        & 371            & $-$6.04  & 0.62\\
\hline
\hline
\end{tabular}
\caption{Wind predictions for a 60 \msun\ model with stellar parameters identical
to model series \#10 from Vink et al. (2000,2001).}
\label{tab:results2}
\end{table}

\begin{figure}
\includegraphics[width=\columnwidth]{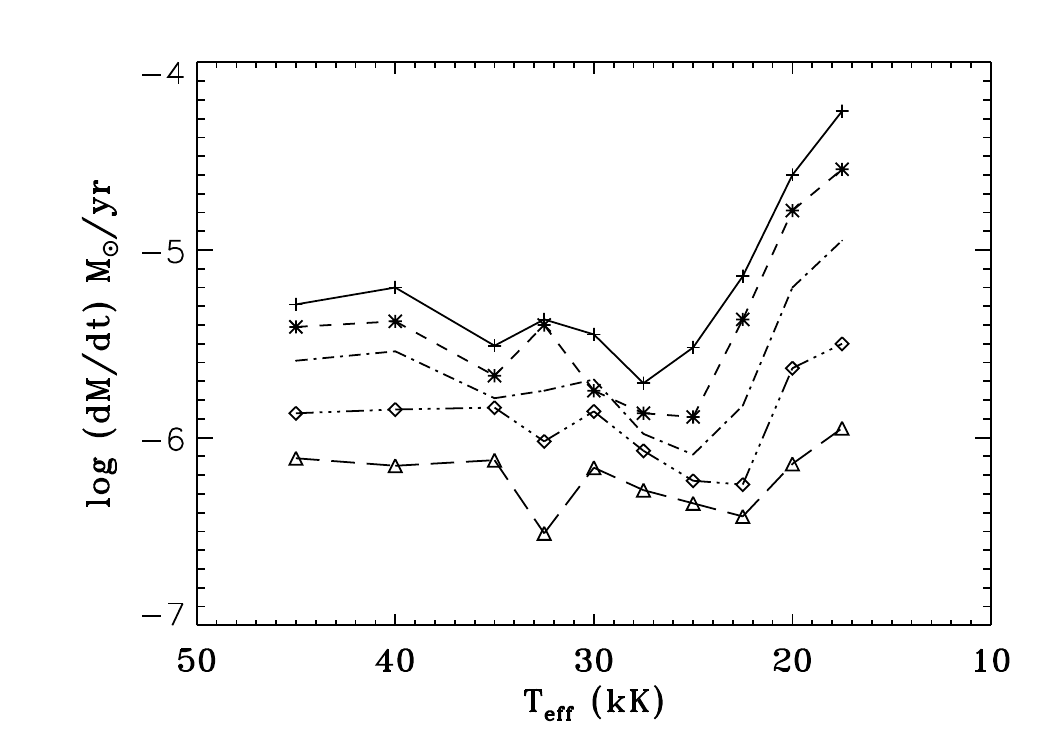}
\caption{The predicted mass-loss rates ($\mdot$) versus \Teff\ for the 60\msun\ star corresponding to series \#10 of Vink et al. (2000, 2001).}
\label{f_mdotM60}
\end{figure}

\begin{figure}
\includegraphics[width=\columnwidth]{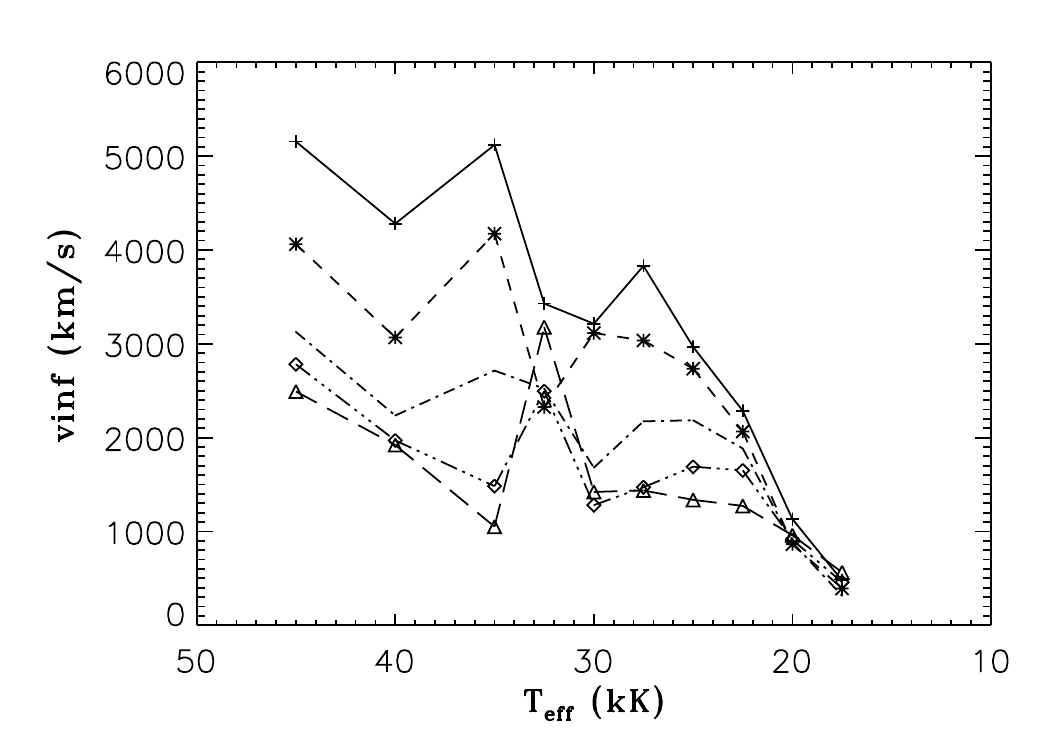}
\caption{The predicted wind velocity ($\vinf$) versus \teff\ for the 60\msun\ star corresponding to series \#10 of Vink et al. (2000, 2001)..}
\label{f_vinfM60}
\end{figure}

\begin{table}
\centering
\begin{tabular}{llllccc}
\hline
\hline
$M$      & $\log L$ & \teff  & $Z/\zsun$ & \vinf  & $\log \mdot$ & \(\beta\)\\
(\Msun)  &  (\lsun)  &  (kK) &           & (km/s) &   (\msunyr) & \\ 
\hline
20      & 5.0     & 45    & 3           & 4090          & $-$6.49  & 0.91\\
        &         &       & 1           & 3149          & $-$6.69  & 0.82\\
        &         &       & 1/3         & 2710          & $-$6.96  & 0.76\\ 
        &         &       & 1/10        & 3078          & $-$7.40  & 0.81\\
        &         &       & 1/33        & 2310          & $-$8.06  & 1.00\\[0.5em]

        &         & 40    & 3           & 4187          & $-$6.71  & 0.88\\
        &         &       & 1           & 2273          & $-$6.62  & 0.87\\
        &         &       & 1/3         & 1733          & $-$6.94  & 0.67\\ 
        &         &       & 1/10        & 1662          & $-$7.28  & 0.64\\
        &         &       & 1/33        & 1568          & $-$7.58  & 0.63\\[0.5em]

        &         & 35    & 3           & 4891          & $-$6.89  & 1.24\\
        &         &       & 1           & 10337         & $-$7.55  & 1.69\\
        &         &       & 1/3         & 6992          & $-$7.79  & 1.51\\[0.5em]

        &         & 32.5  & 3           & 3722          & $-$6.80  & 1.03\\
        &         &       & 1           & 3997          & $-$7.08  & 1.03\\
        &         &       & 1/10        & 4177          & $-$7.91  & 1.12\\
        &         &       & 1/33        & 2204          & $-$8.12  & 0.80\\[0.5em]
    
        &         & 30    & 3           & 4442          & $-$7.04    & 1.28\\
        &         &       & 1           & 4795          & $-$7.20    & 1.04\\
        &         &       & 1/3         & 2571         & $-$7.22    & 0.78\\
        &         &       & 1/10        & 2658          & $-$7.51    & 0.83\\
        &         &       & 1/33        & 2338          & $-$8.00   & 0.87\\[0.5em]

        &         & 27.5  & 3           & 3300          & $-$7.00  & 1.14\\
        &         &       & 1           & 5448         & $-$7.33   & 1.22\\
        &         &       & 1/3         & 2164          & $-$7.25  & 0.77\\
       &         &       & 1/10        &  1526         & $-$7.36   & 0.69\\
        &         &       & 1/33        & 1196         & $-$7.58  & 0.64\\[0.5em]

        &         & 25    & 3           & 3615         & $-$7.10  & 1.15\\
        &         &       & 1           & 2186         & $-$7.10  & 1.02\\
        &         &       & 1/3         & 1757          & $-$7.26  & 0.76\\
        &         &       & 1/10        & 1001          & $-$7.38  & 0.73\\
        &         &       & 1/33        & 678          & $-$7.56  & 0.56\\[0.5em]

        &         & 22.5  & 3           & 2008          & $-$6.37  & 1.02\\
        &         &       & 1           & 1972          & $-$6.90  & 1.01\\
        &         &       & 1/3         & 2053          & $-$7.40   & 0.98\\
        &         &       & 1/10        & 1461          & $-$7.53  & 0.76\\
        &         &       & 1/33        & 838          & $-$7.66  & 0.65\\[0.5em]

        &         & 20    & 3           &  896          & $-$5.92  & 0.76\\
        &         &       & 1           &  1087          & $-$6.33   & 0.83\\
        &         &       & 1/3         & 989          & $-$6.73    & 0.75\\
        &         &       & 1/10        &  903          & $-$7.16    & 0.78\\
        &         &       & 1/33        &  788        & $-$7.68    & 0.73\\[0.5em]

        &         & 17.5   & 1/3         & 500          & $-$6.61   & 0.63\\
        &         &       & 1/10        & 591        & $-$7.09   & 0.67\\
        &         &       & 1/33        & 508          & $-$7.51   & 0.63\\[0.5em]

       &          & 15  & 1/3          & 335          & $-$6.72.  & 0.60\\       
       &         &       & 1/10        & 427           & $-$7.18  & 0.63\\
        &         &       & 1/33        &  467          & $-$7.89  & 0.61\\[0.5em]
				
        &         & 12.5  & 1/10        &  399           & $-$7.47 & 0.62\\
\hline
\hline
\end{tabular}
\caption{Wind predictions for a 20 \msun\ model with stellar parameters identical
to model series \#3 from Vink et al. (2000,2001).}
\label{tab:results3}
\end{table}

\begin{figure}
\includegraphics[width=\columnwidth]{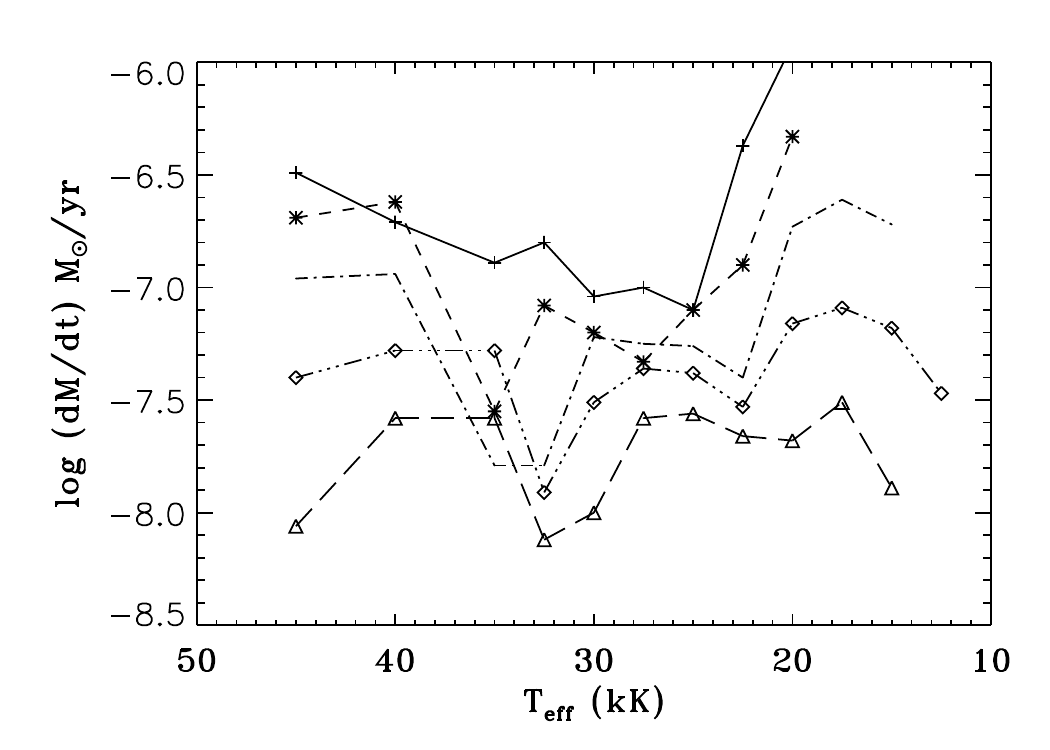}
\caption{The predicted mass-loss rates ($\mdot$) versus \Teff\ for the 20\msun\ star (series \#3) of Vink et al. (2000, 2001).}
\label{f_mdotM60}
\end{figure}

\begin{figure}
\includegraphics[width=\columnwidth]{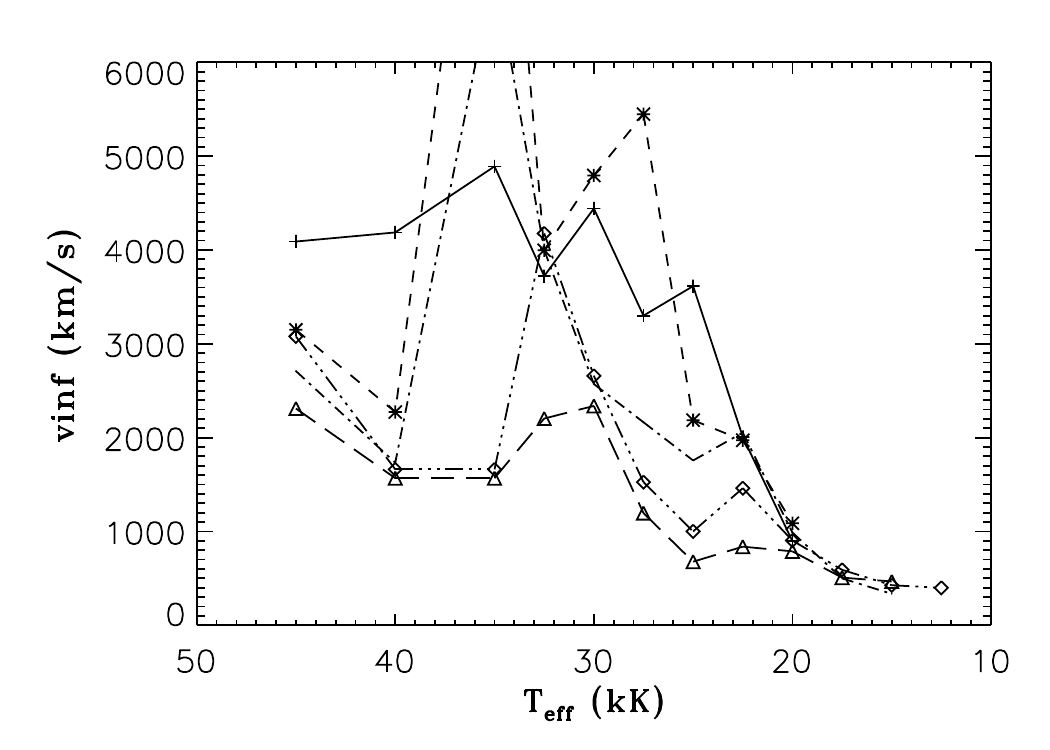}
\caption{The predicted wind velocity ($\vinf$) versus \teff\ for the 20\msun\ star (series \#3) of Vink et al. (2000, 2001).}
\label{f_vinfM20}
\end{figure}

\section{Spectral synthesis of the `weak wind'-bump}

\begin{figure*}
\includegraphics[width=\textwidth]{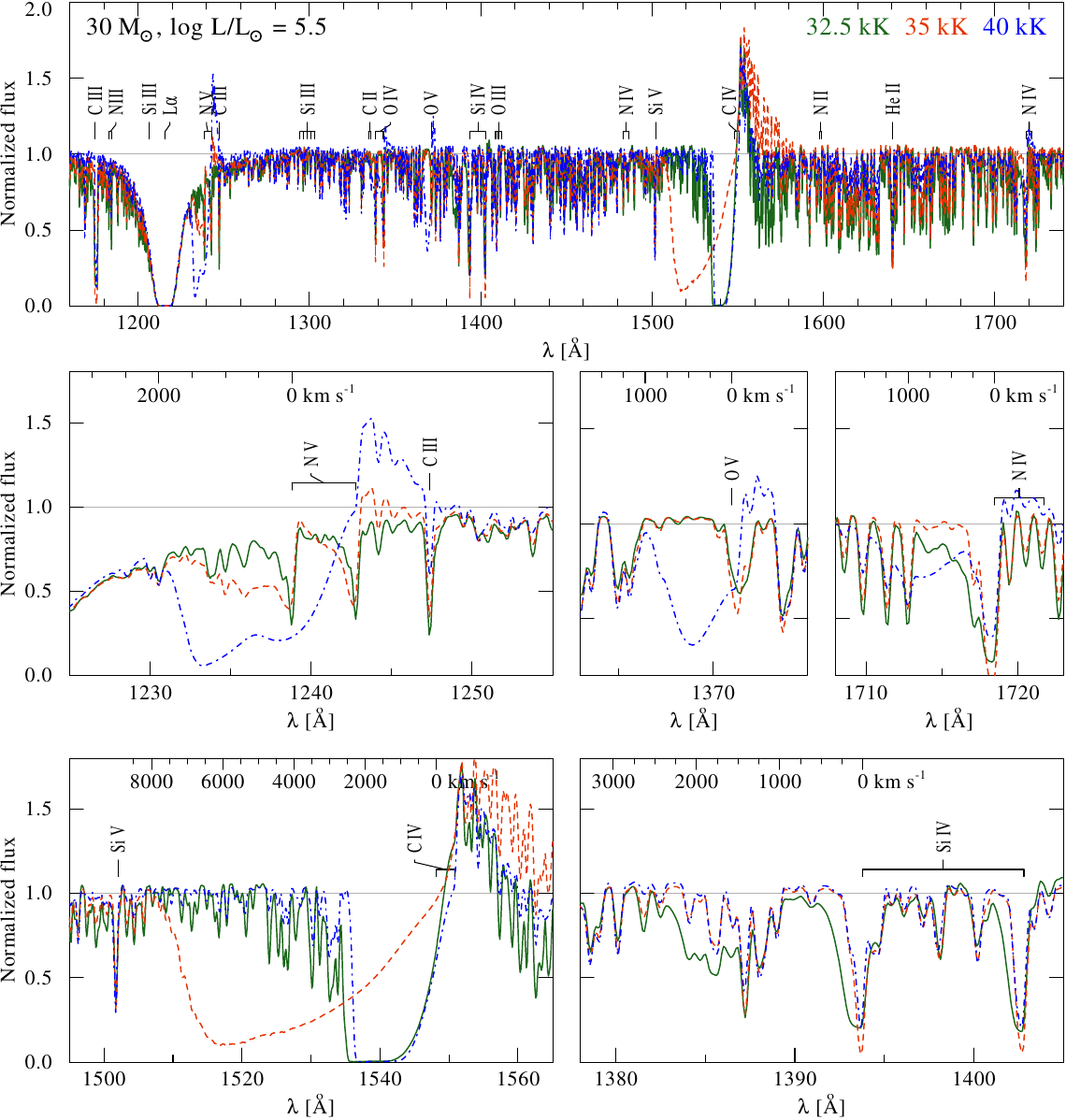}
\caption{The predicted spectral appearance of the $30\,M_\odot$ models with temperatures around $35\,000\,$K at $Z_\odot/3$.
The uppermost panels shows the overall spectrum in the far UV while the smaller panels show zoom-ins around important lines affected by the stellar wind.}
\label{f_specM30}
\end{figure*}

\begin{figure*}
\includegraphics[width=\textwidth]{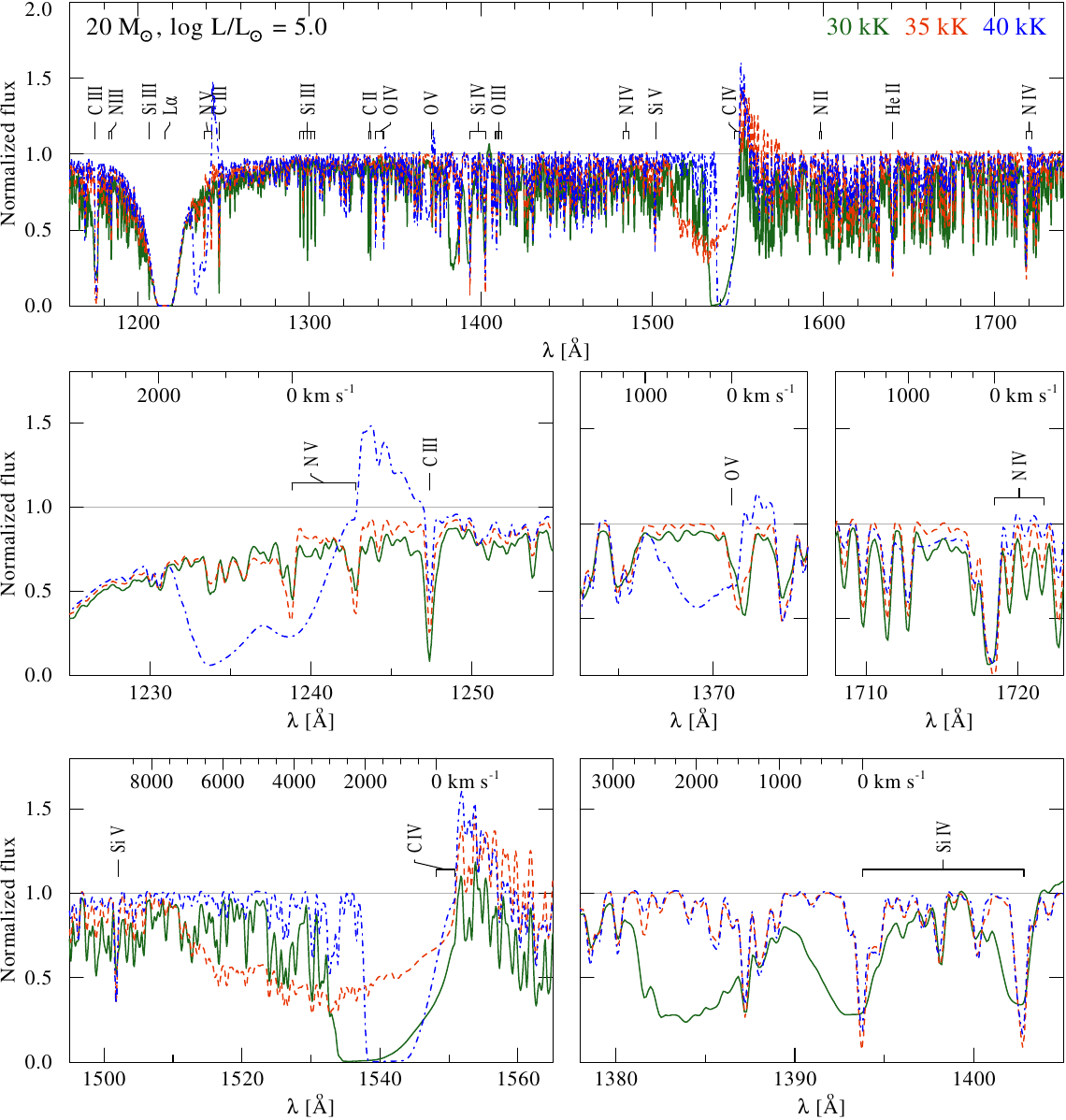}
\caption{The predicted spectral appearance of the $20\,M_\odot$ models with temperatures around $35\,000\,$K at $Z_\odot/3$.
  The uppermost panels shows the overall spectrum in the far UV while the smaller panels show zoom-ins around important lines affected by the stellar wind.}
\label{f_specM20}
\end{figure*}

Over the last few decades it has been established that the UV morphology of OB stars generally forms a sequence in terms of the optically determined stellar effective temperature (e.g. Walborn et al. 1985), and that this is only expected to break around the bi-stability jump at spectral type B1 (Lamers et al. 1995; Vink et al. 1999; Trundle et al. 2004; Crowther et al. 2006; Benaglia et al. 2007; Markova \& Puls 2008; Groh \& Vink 2011). 
It should however be noted that there are indications from time-dependent stellar structure computations that certain temperature ranges might not have the required opacity to launch a stellar wind at each and every location in the HR diagram (e.g. Grassitelli et al. 2020). These opacity gaps may be related to the problem of initiating stellar winds around 35\,000 K, and are perhaps related to the issue of weak winds. 

As discussed in Vink et al. (2009) there is a difference between the weak-wind problem that generally occurs for stars with luminosities below $\log{L}/\lsun < 5.2$ (Martins et al. 2005) and stars that have weak winds {\it for their spectral types} (Walborn et al. 1985). How these latter weak-winded stars are related to the weak-wind problem, and/or OVz stars, and/or whether OVz stars have weaker winds -- due to stronger gravity -- remains an area of study (Sabin-Sanjulian et al. 2014). 

Muijres et al. (2012) suggested that the lower mass-loss rates around 35\,kK might be unphysical and related to the weak-wind regime.
The lack of driving at these temperatures might either result in a complete absence of stellar winds and associated P\,Cygni profiles,
or conversely nature might find alternative ways to launch winds through the the transonic region and reach infinity. In that case, it would
be interesting to see if winds around this temperature regime have canonical wind velocities and mass-loss rates, or if the winds are faster with lower inferred mass-loss rates. Note that this weak-wind experiment may also provide information as to whether the lack of driving below the sonic
point in most CMF approach models is either realised in Nature, an artefact of the modelling approach, or whether additional physics may need to be considered for
pushing winds through the transsonic region.

To investigate whether the high terminal velocities found around $35\,$kK would potentially be observable
in upcoming UV ULLYSES observations, we calculated model atmospheres with the Potsdam Wolf-Rayet (PoWR) model atmosphere
code (e.g. Gr{\"a}fener et al. 2002; Hamann \& Gr{\"a}fener 2003; Sander et al. 2015)
to obtain synthetic spectra in the UV range. PoWR is a non-LTE stellar atmosphere code assuming a spherically symmetric star with a stationary outflow.
The radiative transfer is treated in the co-moving frame. For standard models, the velocity field in the supersonic wind is prescribed
by a $\beta$-type velocity law, while the stratification in the subsonic part follows from a consistent solution of the hydrodynamic equation.
As it is our aim to simulate the spectral appearance of the solutions resulting from Monte Carlo calculations, we treat the mass-loss
rate $\dot{M}$ as well as the wind parameters $\beta$ and $\varv_\infty$ as fixed input parameters instead of using the option to derive
them (as e.g. done in Sander et al. 2017, 2018). 

Starting from pre-calculated grids for OB-type stars (Hainich et al. 2019), we constructed models for $20\,M_\odot$ and $30\,M_\odot$ with
abundances matching those of the Monte Carlo models for $1/3\,Z_\odot$ (with solar values from Anders \& Grevesse 1989). To allow for maximum
alignment, we switch off clumping, use the derived $\dot{M}$-, $\varv_\infty$-, and best-$\beta$-values (cf.\ Tables \ref{tab:results1}
and \ref{tab:results3}) as input, and connect the wind velocity law at a fixed transition velocity of $0.95\varv_\mathrm{sound}$ instead of
using the standard option with a smooth $\varv$-gradient connection. In the spectral synthesis, we assume a depth-dependent
micro-turbulence $\xi$ (Shenar et al. 2015), starting from a minimum photospheric value $\xi_\mathrm{min} = 14\,\mathrm{km}\,\mathrm{s}^{-1}$.
In the wind, $\xi$ scales with the wind velocity as $\xi(r) = 0.1 \varv(r)$.
For our example plots, we further select a rotational broadening of $50\,\mathrm{km}\,\mathrm{s}^{-1}$.

In Figs.\,\ref{f_specM30} and \ref{f_specM20} we depict the predicted far-UV spectra for selected models of
$30\,M_\odot$ and $20\,M_\odot$ around $35\,000\,$K where our Monte Carlo calculations yield significant `dips' in the mass-loss rate.
For both masses, the high terminal velocities accompanying these lower $\dot{M}$-values are only potentially measurable in the
prominent \ion{C}{iv}\,1550\,\AA-doublet, while the other lines at this temperature do not show substantial contributions from the
high-velocity part of the wind. This is different for the higher and lower temperatures, illustrating that the lower mass-loss rates
essentially remove most of the signatures of a very fast wind.

While our calculations are performed for an elemental composition
reflecting scaled solar abundances, empirical analyses of massive stars
often yield enhanced nitrogen abundances. This enhancement is a
well-known side-effect from the CNO cycle converting most of the total
CNO abundance into nitrogen. To test whether such a conversion of the
CNO abundances would effect our diagnostics, we also computed PoWR models
reflecting these `CNO-mixed' abundances. For our interesting
metallicity of $Z_\odot/3$, we investigate the situation of a full
mixture assuming $X_\mathrm{N} \approx 60 X_\mathrm{C} = 60 X_\mathrm{O}$,
yielding mass fractions of $X_\mathrm{N} = 4.43 \cdot 10^{-3}$ and
$X_\mathrm{C} = X_\mathrm{O} = 7.35 10^{-5}$.

\begin{figure}
\includegraphics[width=\columnwidth]{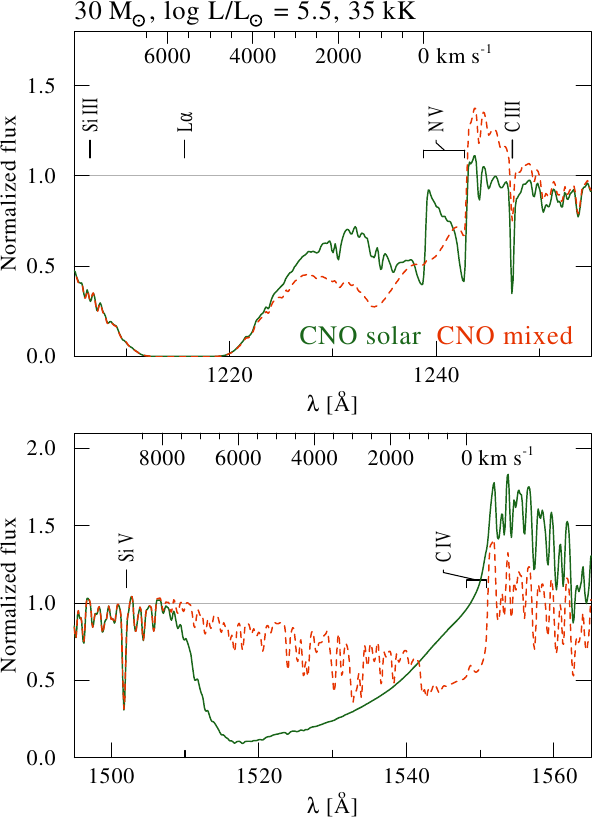}
\caption{The predicted spectral appearance of the $30\,M_\odot$ models with temperatures around $35\,000\,$K at $Z_\odot/3$.
The upper panel shows the \ion{N}{v}\,1250\,\AA-doublet and the lower panel shows the \ion{C}{iv}\,1550\,\AA-doublet.}
\label{f_specM30-cnomix}
\end{figure}

\begin{figure}
\includegraphics[width=\columnwidth]{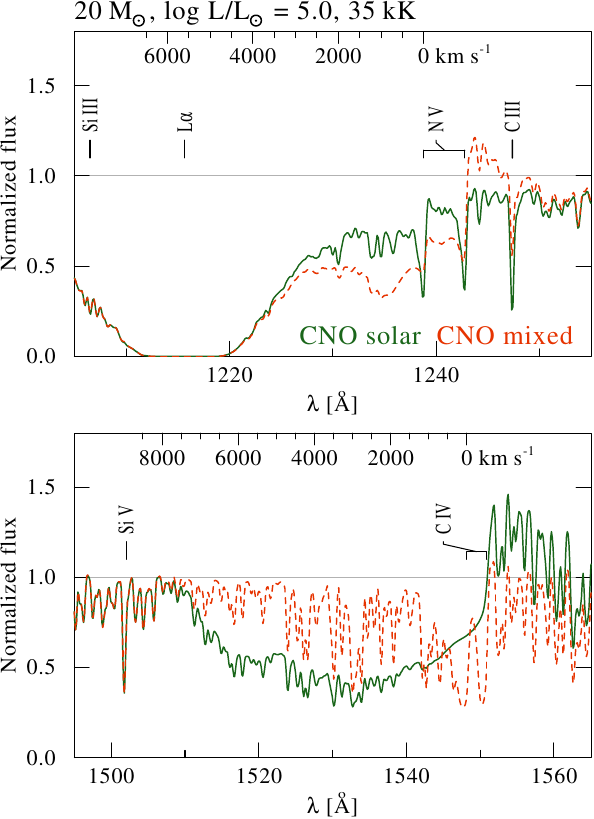}
\caption{The predicted spectral appearance of the $20\,M_\odot$ models with temperatures around $35\,000\,$K at $Z_\odot/3$.
The upper panel shows the \ion{N}{v}\,1250\,\AA-doublet and the lower panel shows the \ion{C}{iv}\,1550\,\AA-doublet.}
\label{f_specM20-cnomix}
\end{figure}

As we illustrate in Figs.\,\ref{f_specM30-cnomix} and\,\ref{f_specM20-cnomix},
a strong depletion of carbon immediately removes
any signature of the high terminal velocity from the
\ion{C}{iv}\,1550\,\AA-doublet. Instead, the
\ion{N}{v}\,1250\,\AA-doublet now shows a pronounced wind profile. While
this means that we would expect a considerable indication of a high
velocity in the UV spectrum also in the case of nitrogen enhancement,
one has to be aware that the interstellar L$\alpha$-absorption would
likely spoil precise measurements for $\varv_{\infty} >
5000\,\mathrm{km}\,\mathrm{s}^{-1}$. This is illustrated in our Figs.\,\ref{f_specM30} to\,\ref{f_specM20-cnomix}, 
where we assumed a typical SMC H-column density taken from a current study of OB stars in NGC\,346 (Rickard et al., in prep). 
According to our test models,
measurements of $\varv_\infty$ will further be complicated due to the
absence of a clear absorption trough in the \ion{N}{v}-line, thereby
blending the high-velocity part of the line profile with the iron forest
and making measurements very susceptible to the S/N-level. However, the \ion{N}{v}-doublet is sensitive to X-rays. Thus, wind-embedded shocks causing Auger ionization (Cassinelli \& Olson 1979) could enhance the \ion{N}{v}-population and consequentially the absorption trough.

\section{Discussion}

Similar to our earlier results in Muijres et al. (2012) and Vink (2018b), the new mass-loss rates are somewhat lower than those
from the Vink et al. (2000, 2001) recipe for an assumed ratio of $\vinf/\varv_{\rm esc} = 2.6$ for O-type stars (Lamers et al. 1995).
However, also similar to M\"uller \& Vink (2008) and these follow-up studies, we overpredict the terminal wind velocities.
Similar overpredictions are well-documented also for modified CAK theory (Pauldrach et al. 1986), and similar high velocities are obtained from
hydro-dynamically consistent CMF
computations (Bj{\"o}rklund et al. 2020).
When we adopt the newly predicted wind velocities as input in the Vink et al. (2000) recipe, the mass-loss rates
are invariant. 

This invariance is not surprising in light of the results obtained for
the wind efficiency $\eta = \dot{M}\varv_\infty / (L/c) $ in Vink et al.
(2000). Vink et al. (2000) demonstrated that the product of
$\dot{M}\varv_\infty$ is approximately constant for constant mass and
luminosity but different assumptions of
$\varv_\infty/\varv_\mathrm{esc}$. Moreover, the modified wind momentum $D_{\rm mom}$ (Puls et al. 1996; Kudritzki et al. 1999), where $\dot{M}\varv_\infty \sqrt{R_\ast/R_\odot}$ scales with $L$ for a constant CAK parameter of 0.6 (in which case the stellar Mass cancels) has been established. Our new results are exactly in line with this presumed scaling of $\dot{M} \propto \varv_\infty^{-1}$, but highlight that a local treatment of the hydrodynamics yields a different parameter combination along this relation than the global approach.

\begin{figure}
\includegraphics[width=\columnwidth]{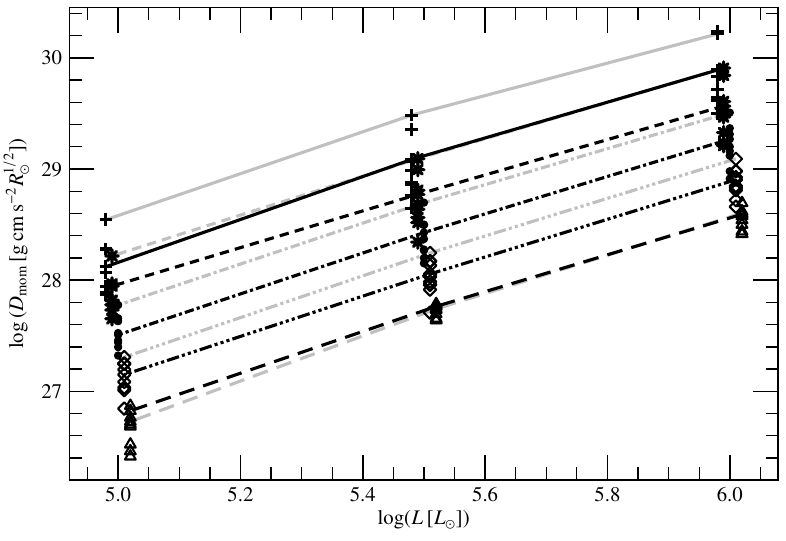}
\caption{The Modified wind momentum-luminosity relation as defined by Puls et al. (1996) for model results. At each $Z$, the different model temperatures have a somewhat different modified wind momentum. To illustrate the effect, all five $Z$-values have 2 lines: the bolder one is for \teff\ $=$ 40\,000 K, while the lighter one is for 20\,000 K. At low $Z$ the bolder and lighter lines overlap, while there is a substantial difference at high $Z$, due to the larger opacities at higher $Z$. To better distinguish the datapoints at each $L$, symbols referring to
the same $Z$ have the same minute offset in the horizontal direction.}
\label{dmom}
\end{figure}

We highlight this in the modified wind momentum plot of Fig.\,\ref{dmom}.
At each $Z$, the different model effective temperatures have a somewhat different modified wind momentum. To illustrate the effect all five $Z$-values have two lines, a
bolder one for \teff\ $=$ 40\,000 K and a lighter one for 20\,000 K. Their purpose is to 
indicate overall trends.
While there are just three steps in $L$, there are many \teff\ datapoints for
each $L$, and it is clear there is significant scatter in $D_{\rm mom}$
for each $Z$ at each $L$. Yet, the reference lines follow a common trend 
in $D_{\rm mom}$ with luminosity. This indicates that it is the actual wind
momentum (i.e. $\dot{M} \times \vinf$) that is responsible for the trends, and it is not the result of  
radius modification due to varying \teff. 

In other words, the 
pronounced temperature-dependence between the 40 kK an d 20kK (bold and lighter lines) does {\it not} stem from differences in radii (which should have been corrected by the 
$\sqrt{R}$ modification of $D_{\rm mom}$), but from true differences in the line
opacities at different \teff. 
This can also be noted from the strong differences between the growing differences in relations with $Z$. At low $Z$, the size of the bi-stability jump is significantly diminished (Vink 2018b) and for similar reasons the offsets between $D_{\rm mom}$ at 20 kK and 40 kK are almost negligible at low $Z$, while they are significant at high $Z$ where the Fe line opacities are more prevalent. This analysis questions the use of $D_{\rm mom}$ for various $Z$ (cf. Mokiem et al. 2007).

The most significant difference with respect to the Vink et al. (2000, 2001) study is the 
shallower mass-loss rate dependence on metallicity of the new dynamically consistent
computations. While the shallower slope of these new mass-loss predictions of \mdot\ vs. $Z^{0.42 \pm, 0.03}$ for stars above the bi-stability jump is lower than the values
  determined empirically of \mdot\ vs. $Z^{0.83 \pm 0.16}$ by Mokiem et al. (2007) and \mdot\ vs. $Z^{2}$ by Ramachandran et al. (2019) for the Large and Small Magellanic Clouds at resp. 0.5\,$\zsun$ and 0.2\,$\zsun$, they may
be in better agreement with the surprising results of Tramper et al. (2014) who found larger than expected mass-loss rates
for the very metal-poor galaxies IC 1613, WLM, and NGC 3109, that have oxygen (O) abundances that are about 10\% solar (Lee et al. 2003; Bresolin et al. 2007). Note however
that this mismatch of H$\alpha$ empirical mass-loss rates in these very low $Z$ galaxies by Tramper et al. and the radiation-driven
wind theory has been challenged by Bouret et al. (2015). An in-depth analysis of these very low-$Z$ O-type stars
with upcoming ULLYSES and recently secured complementary OIR X-Shooter spectra is expected to shed more light on these issues.

Regarding our predicted terminal wind velocities, the observed drop of \vinf\ towards lower \teff\ is in line with observations
(see Garcia et al. 2014 and references therein).
We also predict a shallow but notable dependence of wind velocity on metallicity as $\vinf \propto Z^{0.19}$.
This will become testable against hundreds of O-star spectra in the LMC and SMC with the upcoming  ULLYSES \& X-Shooter data.
Garcia et al. (2014) have shown the complications of their $Z$ dependence from the empirical side, and also showed that the
Fe abundance in these very low $Z$ galaxies mentioned above is larger than expected on the basis of the nebular O abundance, and
in fact more similar to the Fe abundance of the SMC (see also Tramper et al. 2014).
In other words, empirical studies appear to show a {\it sub}solar [$\alpha$/Fe] ratio, which may be counter-intuitive considering that
lower $Z$ low-mass stars in the Milky Way show enhanced [$\alpha$/Fe] ratio by approx. 0.4 dex at low $Z$ (see Vink et al. 2001 and
references therein). Non-solar metallicity scaled [$\alpha$/Fe] conversions as a function of metallicity can be found in Table 5 of Vink et al. (2001).

We next turn our attention to the capricious behaviour of wind properties in the the range around 35\,000 K. We earlier attributed this
to the same issues Muijres et al. (2012) experienced for lower mass/luminosity models, where this was attributed
to the onset of the "weak-wind problem" (Martins et al. 2005; Puls et al. 2008; Marcolino et al. 2009) that might be related to the `inverse' bi-stability effect
due to a lack of Fe {\sc iv} driving around spectral type O6.5 -- corresponding to $\log(L/\lsun) = 5.2$.
If the `weak wind problem' is not only directly related to low L, but also the specific Teff of approximately 35000 K [corresponding to spectral type O6.5] we
may possibly expect objects of order 40 000 K and higher to again show stronger wind features, even when the objects remain below l$\log(L/\lsun) = 5.2$.
This should be testable with the new ULLYSES observations for O-type stars in this temperature range.
It may also be feasible to be able to witness extremely fast winds with $\vinf > 5\,000$ km/s for the weak wind stars around 35\,000 K.
A similar finding of a `weak wind'-bump was also found for supergiants at SMC metallicity in the recent CMF-based calculations by Bj{\"o}rklund et al. (2020). 
 The presence or absence of such a regime with weaker, but faster winds in the in large observational samples such as X-Shooting ULLYSES (XSHOOTU) will mark an indicator for all
  current mass-loss prediction efforts that could provide important hints on whether or not additional wind driving physics need to be considered in order to explain the empirical results.

\section{Conclusions}

We have computed a new set of dynamically consistent mass-loss rates and terminal wind velocities spanning the range of the O-type stars and the B supergiants, as a function
of metallicity, and we present a new mass-loss recipe in the Appendix.
We generally find wind velocities for the O-type stars that are larger than observed empirically, which also means that the predicted mass-loss
rates are slightly smaller for these larger wind velocities than applying a fixed observed ratio of $\vinf/\varv_{\rm esc} = 2.6$ for O-type stars as was adopted in the Vink et al. (2000, 2001)
  recipe. 

We furthermore uncover a dichotomy of the O versus B supergiants across the bi-stability jump with respect to the $Z$-dependence of the wind properties.
  The O-type stars showcase a dependence of the wind velocity on $Z$ that is inline with earlier modelling of Leitherer et al. (1992), leading to a shallower dependence of the
  mass-loss rate on $Z$ as was found in the global wind models of Vink et al. (2001). However, the B supergiants do not show a dependence of wind velocity on $Z$ (nor $M$) and
  the $Z$-dependence of $\dot{M}$ on $Z$ with respect to Vink et al. (2001) is unaffected for this temperature range.

\section*{Acknowledgements}

J.S.V. and A.A.C.S. are supported by STFC funding under grant number ST/R000565/1. 
The spectral synthesis figures in this work were created with \textsc{WRplot}, developed by W.-R.\ Hamann. We thank an anonymous referee for providing constructive comments that helped improve the paper.

\section*{Data availability}
 
The data underlying this article will be shared on reasonable request to the corresponding author.

\appendix
\section{Python mass-loss recipe}

The accompanying Python mass-loss recipe can be downloaded from https://armagh.space/jvink, or is available upon request.

Here follow some explanatory notes on the usage of the mass-loss recipe.
We envision the mass-loss recipe to be used for at least two purposes. The first one is for comparison with observations and alternative mass-loss predictions, the second one is for utilisation in stellar evolution models, in particular where it relates to rapid changes at the first and second bi-stability jumps.

In comparison to the earlier Vink et al. (2000) computations, the current models are dynamically consistent, which has resulted in larger terminal wind velocities than were adopted from observed Lamers et al. (1995) wind velocity relations in the original 2000 recipe. (Note that similar over predictions are common in modified CAK theory and in recent co-moving frame (CMF) computations in Bj\"orklund et al. 2000). This different treatment of the wind velocities has an impact on the characteristic wind densities, which are utilised in the recipes (both original IDL script, and current Python script) to place objects on different sides of the bi-stability jumps. While the new Monte Carlo computations place the first bi-stability jump at a lower \teff\ of 21 kK instead of approx 25 kK, the implementation of the characteristic density should not be replaced for reasons of internal consistency.  

As discussed in Vink et al. (2000), placing observed objects on the correct side of the bi-stability jump(s) should always take precedence over the utilisation of the characteristic wind density tools underlying Eqs. (4) and (5) in Vink et al. (2000), which can be wild extrapolations. 
As was also discussed in Sect. 5.3 in Vink et al. (2000), predictions on the cool side of the second bi-stability jump are highly speculative, as the characteristic density for the second jump is very rough as it is based on just a few low \teff\ models, and the implementation of this second bi-stability jump in for instance stellar evolution models should only be performed with caution. 

Given the fact that Lamers et al. (1995) noted a second bi-stability jump at spectral type A0 (\teff\ approx 10 kK) Vink et al. (2000) provided the option to include this in stellar evolution modelling. It was considered more physically inaccurate to {\it not} include the evidence for the second bi-stability jump than to include it. 
However, in actual stellar evolution modelling, choices have to be made as to where to switch between hot-star and cool-star prescriptions. In Brott et al. (2011) our knowledge of the physics of the second bi-stability jump was utilised to gradually transition from a hot to cool star prescription at approx 10 kK. Also in the oft-employed "Dutch" mass loss prescription in MESA this switch occurs at a fixed \teff\ of 10 kK rather than an extrapolation of the second bi-stability jump to higher \teff. 

Note that these types of evaluations depend on the specifics of the stellar tracks. 
When stars simply evolve from the blue to the red part of the HR diagram, the exact \teff\ location of the bi-stability jumps is basically irrelevant. The key issue is whether these absolute increases in mass-loss rates at the bi-stability jumps are physically correct in the first place. 

However, when stars hover around a certain \teff\ location, the exact location of the bi-stability jumps is absolutely critical (see the discussion in Vink et al. 2020).

\bsp	
\label{lastpage}
\end{document}